# Geometry-Driven Passive Fluid Transport in Paper-Based Microdevices

Mohd Shafiq Nasir[1,*], Akimitsu Kugimiya[1], Muhammad Salman Al Farisi[1,2,*]

[1]*Department of Biomedical Information Sciences, Hiroshima City University, Hiroshima 731-3194, Japan*

[2]*Graduate School of Life Science and Systems Engineering, Kyushu Institute of Technology, Kitakyushu 808-0196, Japan*

* Correspondence: dj65002@e.hiroshima-cu.ac.jp (MSN), alfarisi-m788@m.kyutech.ac.jp (MSAF)

**Abstract**

Channel geometry strongly influences capillary-driven fluid transport in paper-based devices, yet systematic comparative studies correlating geometric design with flow behaviour and analyte confinement remain limited. The present study investigates five distinct channel geometries namely converging-diverging, diverging-converging, wide-to-narrow, circular, and rectangular that was fabricated on cellulose filter paper with a standardized area of 32.5 mm² and analyzed using geometry-adapted extensions of the Lucas–Washburn equation. Pyrene and benz[α]anthracene were employed as fluorescent model analytes to enable UV-based quantification of analyte confinement within each geometry. Flow transport times ranged from 23.1 s (circular, fastest) to 65.0 s (diverging-converging, slowest), with corresponding mean velocities of 0.571 and 0.284 mm/s for pyrene respectively, demonstrating that channel geometry strongly influences capillary transport in paper-based devices. Diverging-converging and wide-to-narrow designs produced the greatest analyte confinement by imposing flow retardation and sustained channel acceleration respectively, while circular and rectangular designs yielded relatively uniform velocity distributions and weaker confinement. Cyclodextrin-functionalized chitosan coatings served as a surface chemistry tool to anchor analyte retention at designated preconcentration zones, enabling geometric effects to be isolated and quantified. Computational fluid dynamics simulations, calibrated against experimental flow data and validated through a mesh independence study, reproduced the experimentally observed velocity magnitude distributions across all five geometries, showing semi-quantitative agreement with geometry-adapted Lucas–Washburn predictions. These findings establish a geometry-based design framework for passive fluid control in paper-based devices, with implications for lab-on-paper systems, point-of-care diagnostics, and passive preconcentration platforms.

## 1. Introduction

Paper-based microdevices have attracted growing interest as low-cost, portable platforms for passive fluid manipulation. Their operation relies on capillary-driven flow through porous cellulose networks, eliminating the need for external pumps and enabling simple device fabrication and operation. These features have made paper-based devices attractive for applications ranging from point-of-care diagnostics and environmental sampling to food safety screening (Kugimiya et al. 2022, 2025; Shahid et al. 2024). A key engineering challenge, however, lies in controlling the flow behavior of fluids within the device: the rate, spatial distribution, and confinement of liquid transport directly determine device performance, and these properties are difficult to tune without modifying the bulk substrate or applying external stimuli.

The Lucas–Washburn (LW) equation provides the classical theoretical foundation for capillary-driven flow in porous media (Kim et al. 2020; Manousi and Zachariadis 2020). In its standard form, the equation describes penetration length as a function of surface tension, viscosity, pore radius, and contact angle, predicting that the fluid front advances with the square root of time. This model, however, assumes a uniform and constant channel cross-section, a condition rarely met in practice when channels are intentionally patterned with varying geometries. Modifications to the LW equation have been proposed to account for converging, diverging, and radial geometries (Xiao et al. 2023), yet systematic experimental validation of these geometry-adapted equations across multiple channel types coupled with computational fluid dynamics (CFD) verification has not been reported for paper-based devices. Understanding how macroscopic channel design superimposes on the intrinsic porous-medium flow described by LW theory is therefore both a fundamental and practically important open question.

Geometric manipulation of flow channels offers a purely passive strategy to modulate capillary pressure, local velocity, and analyte residence time without chemical modification of the bulk substrate. Prior studies have demonstrated that channel width, branching topology, and lateral geometry influence wetting dynamics in paper microdevices (Kim et al. 2020; Jeon et al. 2021; Gharib et al. 2022; Musile et al. 2023). However, these investigations have typically examined isolated geometric features rather than performing controlled, side-by-side comparisons of distinct channel types under identical fabrication conditions and standardized surface area. Furthermore, most such studies have not coupled experimental measurements with CFD simulation validated through mesh independence analysis, leaving a gap in quantitative understanding of how geometry-induced velocity gradients translate to macroscopic fluid behavior.

To systematically characterize geometry-driven flow behavior, model analytes with well-defined physicochemical properties and easily detectable optical signals are required. Pyrene and benz[α]anthracene, two polycyclic aromatic hydrocarbons (PAHs), are well-suited for this role: both are highly hydrophobic, exhibit strong UV fluorescence at 365 nm, and differ in molecular size and hydrophobicity, enabling discrimination of geometry-dependent confinement effects across analyte types (Manousi and Zachariadis 2020; Sousa et al. 2022; Nasir and Kugimiya 2026). These compounds are used here not as detection targets for environmental monitoring, but as fluorescent physicochemical tracers whose spatial distribution under UV light provides a direct readout of analyte confinement resulting from geometric flow control.

To enhance analyte retention at designated zones and thereby amplify the measurable signal of geometric confinement, cyclodextrin-functionalized chitosan coatings were applied to preconcentration zones. Cyclodextrins are cyclic oligosaccharides with a hydrophobic interior cavity capable of forming inclusion complexes with nonpolar molecules such as PAHs (Alsadun et al. 2024; Huang et al. 2024; Fang et al. 2025). Immobilized within a crosslinked chitosan matrix, they act as surface-bound molecular anchors that localize analyte accumulation at defined positions, enabling fluorescence-based quantification of how much analyte each geometry retains. The cyclodextrin coating is therefore a measurement tool providing the retention signal necessary to quantify geometric effects rather than the primary scientific contribution of this study.

In this work, five channel geometries were fabricated on Advantec No. 2 filter paper at identical surface areas and evaluated experimentally for flow transport time and analyte confinement. Geometry-adapted LW equations were used to predict and interpret the flow behavior of each design. Three-dimensional CFD simulations were calibrated against experimental flow data and validated by a mesh independence study. By systematically linking geometric design to theoretical predictions, experimental measurements, and CFD-validated velocity distributions, this study establishes a quantitative framework for rational geometry selection in passive paper-based microdevices.

## 2. Materials and Methods

### *2.1. Reagents and Materials*

Methanol (99.5%, Nacalai Tesque, Japan) was used as the solvent for dissolving the model analytes. Filter paper (Advantec No. 2, Toyo Roshi Kaisha, Japan) served as the porous substrate for device fabrication, selected for its uniform pore structure and reproducible wicking characteristics (Kugimiya et al. 2022, 2023, 2025; Nasir and Kugimiya 2026) . Chitosan (Tokyo Chemical Industry, Japan) was employed as a biopolymer matrix for surface functionalization. Pyrene and benz[α]anthracene (Tokyo Chemical Industry, Japan) were chosen as model fluorescent analytes based on their intrinsic UV fluorescence at 365 nm, high hydrophobicity, and differing molecular dimensions, properties that make them well-suited tracers for quantifying geometry-dependent analyte confinement. α-Cyclodextrin and β-cyclodextrin (FUJIFILM Wako Pure Chemical Corporation, Japan) were used as host molecules to form inclusion complexes with the model analytes, providing a surface retention mechanism that enables spatial localization of analyte accumulation for fluorescence-based quantification. Acetic acid was used to dissolve chitosan; glutaraldehyde (25% solution) served as a crosslinking agent to stabilize the chitosan–cyclodextrin coating. Sodium dodecyl sulfate (SDS) was applied as a washing reagent to remove unbound analyte residues prior to measurement. All chemicals were used as received without further purification.

### *2.2. Geometric Design of Paper-Based Channels*

Five channel geometries were designed with a standardized preconcentration zone surface area of 32.5 $mm^2$ and a substrate thickness of 260 μm (Advantec No. 2 filter paper). The geometry types were selected to represent distinct flow regimes predicted by geometry-adapted LW theory: (1) converging-diverging, (2) diverging-converging, (3) wide-to-narrow, (4) circular, and (5) rectangular. Each geometry was designed to impose a different balance of capillary pressure modulation and velocity distribution, enabling systematic comparison under controlled conditions.

The governing flow equations for each geometry are presented below. The standard LW equation (Eq. 1) describes capillary penetration in a uniform porous channel and provides the baseline. Geometry-adapted forms (Eqs. 2–6) modify this relationship to account for the local cross-sectional variation imposed by each design.

Standard Lucas–Washburn equation for uniform porous channels:

$$L^2 = \frac{\gamma\, r \cos\theta}{2\mu} t \quad \text{(Eq. 1)}$$

Converging channel (local velocity derived from LW capillary pressure balance, along channel axis):

$$v(x) = \frac{\gamma\, h \cos\theta}{6\mu\, x} \quad \text{(Eq. 2)}$$

Diverging channel (local velocity derived from LW capillary pressure balance, as function of position and width increment):

$$v(x) = \frac{\gamma\, h \cos\theta}{6\mu\, (x + \Delta w)} \quad \text{(Eq. 3)}$$

Wide-to-narrow transition (capillary pressure difference between channel sections):

$$\Delta P = \frac{2\gamma \cos\theta}{r_{\mathrm{n}}} - \frac{2\gamma \cos\theta}{r^{\mathrm{W}}} \quad \text{(Eq. 4)}$$

Circular (radial) channel (squared radial fluid front position):

$$r^2(t) = \frac{\gamma\, h \cos\theta}{2\mu} t \quad \text{(Eq. 5)}$$

Rectangular channel:

$$L^2 = \frac{\gamma\, h\, w \cos\theta}{3\mu\, (h + w)} t \quad \text{(Eq. 6)}$$

In these equations, $L$ is the capillary penetration length; $r$ is the effective pore radius of the paper substrate; $r_n$ and $r^W$ are the effective hydraulic radii of the narrow and wide channel sections in the wide-to-narrow design, respectively; $t$ is elapsed time; $\gamma$ is surface tension; $\theta$ is contact angle; $\mu$ is dynamic viscosity; $v(x)$ is local fluid velocity at position $x$ from the inlet; $\Delta w$ is the width increment along the diverging channel; $h$ is paper thickness; $w$ is channel width; $r(t)$ is the radial fluid front position; and $\Delta P$ is the pressure difference across the wide-to-narrow transition. All variables are expressed in SI units. Equations 2 and 3 are derived from the LW capillary pressure-viscous resistance framework following (Berthier et al. 2015, 2016; Cai et al. 2021a, b; Kumar et al. 2024); Eq. 4 follows the capillary pressure approach of (Xiao et al. 2023).

Two-dimensional and three-dimensional channel designs were generated in Autodesk Fusion and physically fabricated by cutting Advantec No. 2 filter paper using a Graphtec cutting plotter (CE6000-40, Graphtec Corporation, Kanagawa, Japan). The geometric profiles of all five designs are summarized in Table 1, and the 2D layouts are illustrated in Figure 1.

Table 1. Geometric profile of the five paper-based channel designs. All designs share a standardized preconcentration zone area of 32.5 $mm^2$.

| **Design** | **Channel Type** | **Surface Area ($mm^2$)** | **Channel Length (mm)** |
|---|---|---|---|
| 1 | Converging-diverging | 32.5 | 18.40 |
| 2 | Diverging-converging | 32.5 | 18.45 |
| 3 | Wide-to-narrow | 32.5 | 18.38 |
| 4 | Circular | 32.5 | 13.21 |
| 5 | Rectangular | 32.5 | 13.46 |

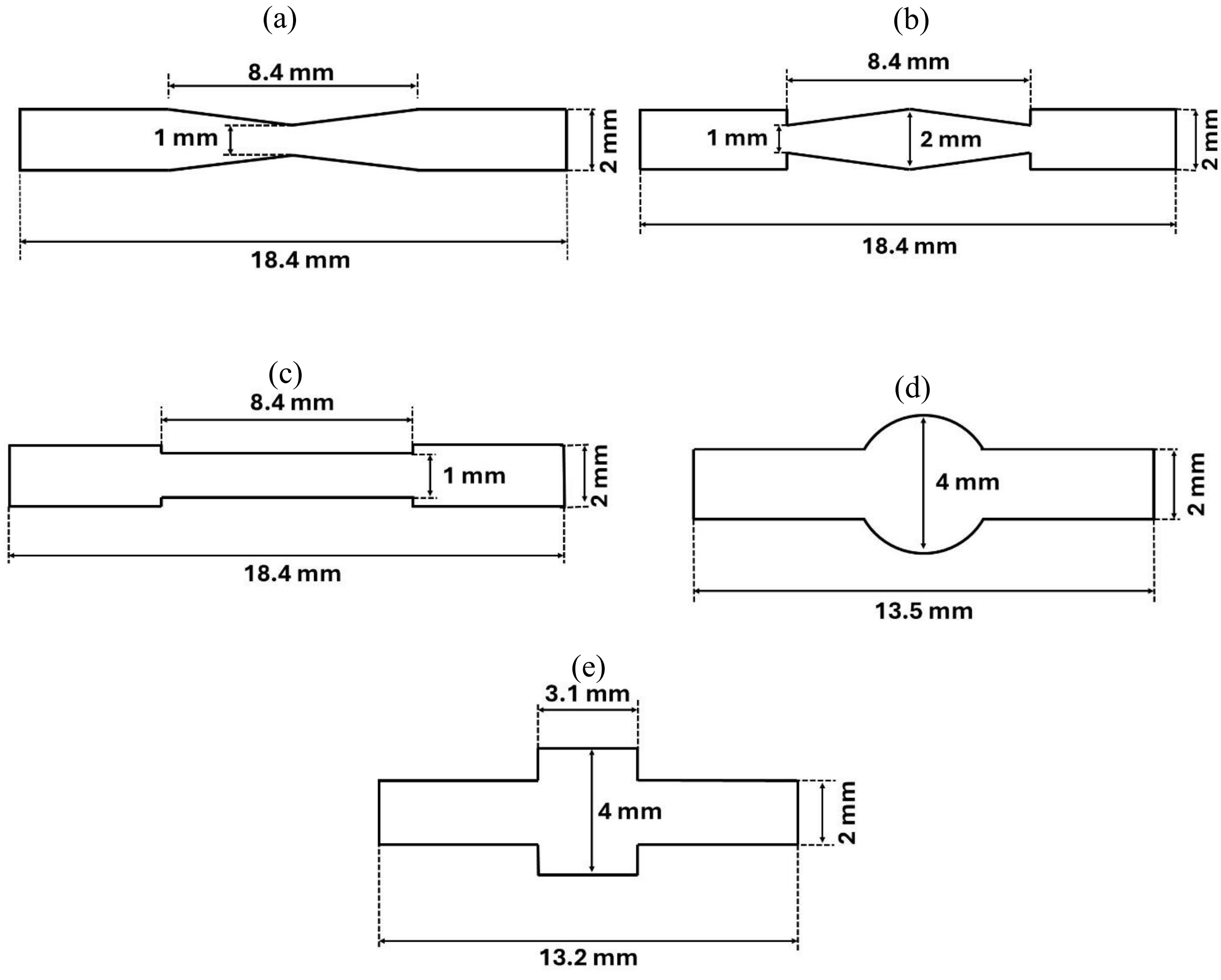

Figure 1. Two-dimensional geometric designs (dimensions in mm): (a) Design 1: converging-diverging; (b) Design 2: diverging-converging; (c) Design 3: wide-to-narrow; (d) Design 4: circular; (e) Design 5: rectangular.

### *2.3. Surface Functionalization of Paper Channels*

To enable spatial quantification of analyte confinement, preconcentration zones within each channel were coated with a cyclodextrin-functionalized chitosan layer. This surface chemistry serves as a molecular anchor, localizing analyte accumulation at defined positions so that fluorescence-based quantification reflects the retention imposed by each geometry. Model analyte solutions were prepared at fixed concentrations (pyrene: 1.2 mM; benz[α]anthracene: 2.5 mM in methanol) to ensure sufficient UV-fluorescence signal for brightness analysis.

A 1:1 (v/v) mixture of chitosan (1% w/v in acetic acid) and cyclodextrin solution was applied to the preconcentration zone (3 μL) and cured by heat-fixing at 80°C for 1 hour, following a surface functionalization approach previously established for PAH detection on Advantec No. 2 filter paper (Nasir and Kugimiya 2026). Devices were then washed sequentially with methanol, 14% SDS, and distilled water to remove unbound analyte and template residues. A final coating of cyclodextrin-chitosan (1:1 v/v) crosslinked with glutaraldehyde (2.5%) was applied and dried at room temperature. Devices were stored in darkness at room temperature prior to use. Control devices without cyclodextrin functionalization were also prepared to isolate the contribution of geometric design alone to analyte confinement.

### 2.4. Evaporation Control Study

To assess whether evaporation contributed independently to the observed differences in flow behaviour across geometries, a gravimetric evaporation control study was conducted. Cyclodextrin-functionalized devices were loaded with 6 μL of pure methanol, replicating the solvent used in analyte experiments. Analyte concentrations in the actual experiment (pyrene: 1.2 mM; benz[α]anthracene: 2.5 mM) are sufficiently dilute that solution viscosity is negligibly different from pure methanol, ensuring that the control accurately represents evaporative conditions during actual assays. Each device was weighed immediately after loading ($m_1$) using an analytical balance and again 30 seconds after the wetting front reached the channel terminus ($m_2$), consistent with the post-flow rest period applied in all device assays (Section 2.4). Evaporative mass loss was calculated as ($m_1 - m_2$) and expressed as the percentage of methanol mass lost by evaporation relative to the initial loaded mass ($m_1$). Five replicates were performed per geometry ($n = 5$).

### *2.5. Flow Characterization and Fluorescence Quantification*

A 6 μL aliquot of model analyte solution was pipetted onto each device inlet and allowed to flow through the channel by capillary action. Flow transport time was recorded from the moment of application until the fluid front reached the channel terminus. Devices were allowed to rest for 30 seconds after flow completion before washing with the SDS solution to remove excess unbound analyte.

Fluorescence imaging was performed under controlled conditions to ensure reproducibility of RGB extraction. The smartphone camera (Apple iPhone 12, 12 MP, aperture f/1.6) was mounted at a fixed distance of 15 cm above the device at 1× magnification (Figure 2). Prior to each image capture, autofocus was locked onto the preconcentration zone and exposure was set to −0.3 EV to prevent signal saturation under UV illumination. ISO was maintained between 1000 and 2000 across all captures. Imaging was conducted in a dark enclosure to eliminate ambient light interference, with a single UV lamp (365 nm) fixed at a consistent position above the device to ensure uniform illumination across all designs. RGB values were extracted from the preconcentration zone using Ryuhou colour picker software on an 8-bit scale (0–255) with background correction performed at the device level using non-functionalized control PADs processed under identical conditions, with selection confined to the confinement zone area for all replicates. All assays were conducted at room temperature in five replicates per geometry per analyte. Images were captured in JPEG format. Brightness was calculated as the mean of the RGB channel values:

$$Brightness = (R_a + G_a + B_a) / 3 \qquad \text{(Eq. 7)}$$

where $R_a$, $G_a$, and $B_a$ are the red, green, and blue channel values extracted from the analyte accumulation zone. This metric provides a reproducible proxy for analyte concentration in the UV-fluorescence images. Brightness values for cyclodextrin-functionalized devices were compared against non-functionalized controls to isolate the geometric contribution to analyte confinement.

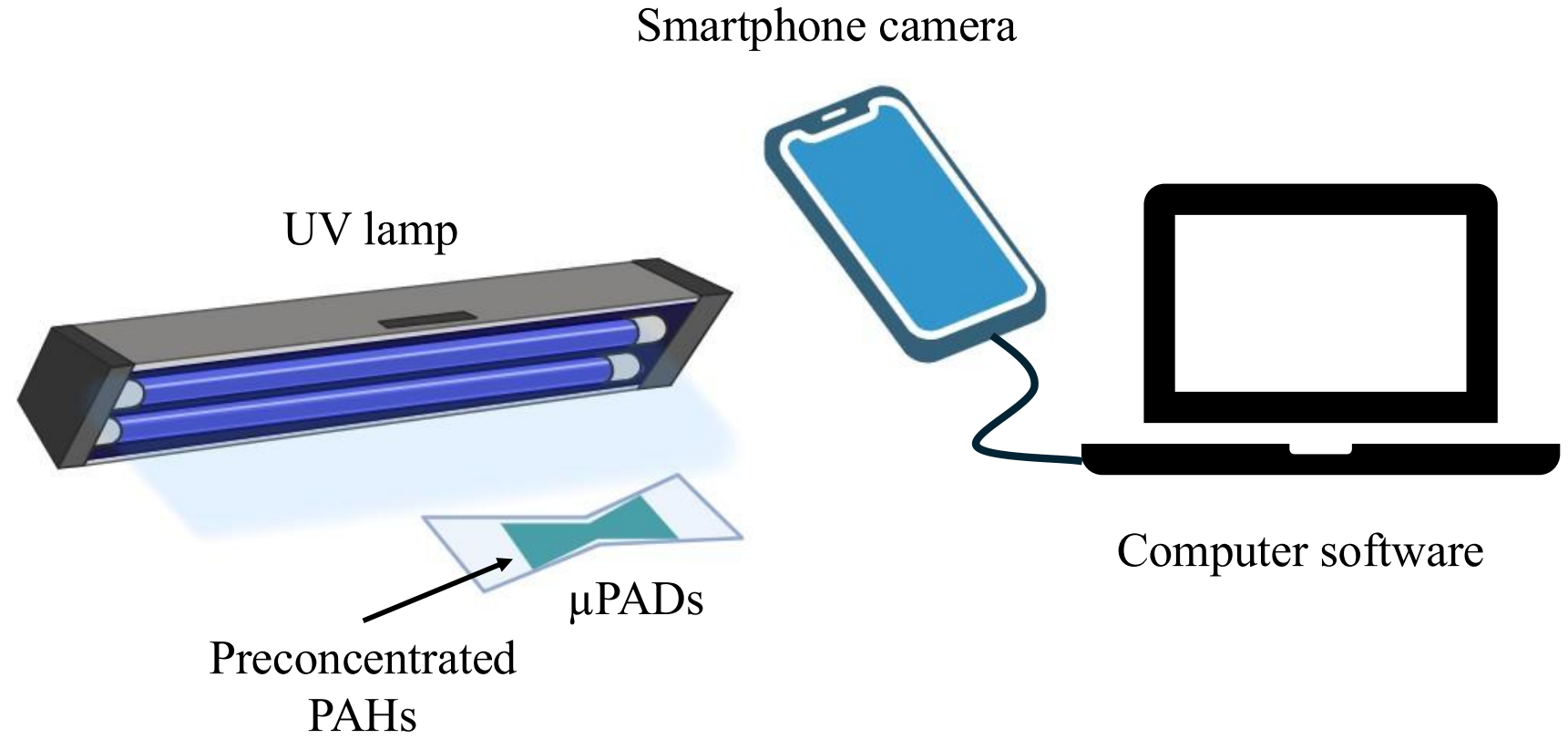

Figure 2. Experimental setup for UV fluorescence visualization of analyte confinement in paper-based channel devices. A UV light source (365 nm) was positioned above the device, and a smartphone camera (Apple iPhone 12) was used to capture fluorescence images of the preconcentration zones. RGB brightness values were subsequently extracted from the captured images using Ryuhou color picker software.

*2.6. CFD Simulation*

Three-dimensional representations of all five channel geometries were imported into Autodesk CFD software for velocity distribution analysis. The goal of the simulations was to validate that the velocity patterns predicted by geometry-adapted LW theory (Eqs. 1–6) are reproduced computationally, providing a CFD-based confirmation of the experimental flow behavior.

Due to the limitations of the software in modeling porous media at the pore scale, the paper substrate was represented as a simplified hollow microchannel with fluid material properties assigned throughout the channel domain, and the surrounding structural boundary defined as a solid wall. This approach captures macroscopic velocity gradients induced by channel geometry while abstracting pore-level effects. Inlet velocity boundary conditions were derived from experimentally measured wetting-front progression rates for each design, ensuring that the simulated flow rate matched the macroscopic behavior of the physical device. Outlet gauge pressure was set to zero to represent open-atmosphere exit conditions. Simulations were run for 500 iterations, and velocity magnitude distributions were recorded with high-velocity regions shown in yellow-to-red and low-velocity regions in green-to-blue.

A mesh independence study was conducted prior to final simulations for all five designs. Three mesh refinement levels were evaluated: coarse (~5,100–7,900 elements), medium (10,000–17,000 elements), and fine (100,000–18,000,000 elements). Mesh independence was confirmed when the difference in outlet velocity between the coarse and medium mesh levels was below 3%. The lowest medium-size mesh satisfying this criterion was selected for each geometry, balancing computational efficiency with numerical accuracy.

## 3. Results and Discussion

### *3.1. Effect of Channel Geometry on Capillary Flow Transport*

The present study examines the influence of channel geometry on capillary flow transport through its flow transport time, CFD-derived velocity magnitude distributions, mean velocity comparison against experimental data, and its relation to the LW equations. It is noted that the five channel designs differ not only in geometry but also in channel length, as a consequence of maintaining a fixed preconcentration zone surface area of 32.5 mm² across all designs (Table 1). Under this constraint, channel length is an intrinsic property of each geometric form rather than an independently controlled variable, since a circular geometry of fixed area necessarily produces a shorter flow path than a diverging-converging geometry of the same area. To assess whether channel length rather than geometry drives the observed transport time differences, designs of comparable channel length were examined. Designs 1, 2, and 3 share nearly identical channel lengths (18.38–18.45 mm) yet exhibit transport times ranging from 32.9 s to 65.0 s for pyrene, a difference of approximately 97%, confirming that geometry rather than channel length is the primary determinant of transport time within this group. The shorter channel lengths of Designs 4 and 5 (13.21 and 13.46 mm respectively) contribute to their faster transport, but the velocity data in Table 2 further confirms that these designs sustain higher mean wicking velocities, consistent with their geometric flow characteristics rather than length alone.

Figure 3 presents the flow transport times for pyrene and benz[α]anthracene across all five channel designs. All geometries completed fluid transport passively within 65 seconds, confirming that capillary-driven flow is sustained across all designs without external pressure assistance. Despite a fixed surface area across all designs, meaningful variation in complete transport time was recorded, demonstrating that channel geometry strongly influences flow speed in paper-based devices. Design 2 produced the slowest transport at 65.0 s for pyrene and 64.9 s for benz[α]anthracene, while Design 4 recorded the fastest at 23.1 s and 31.9 s respectively.

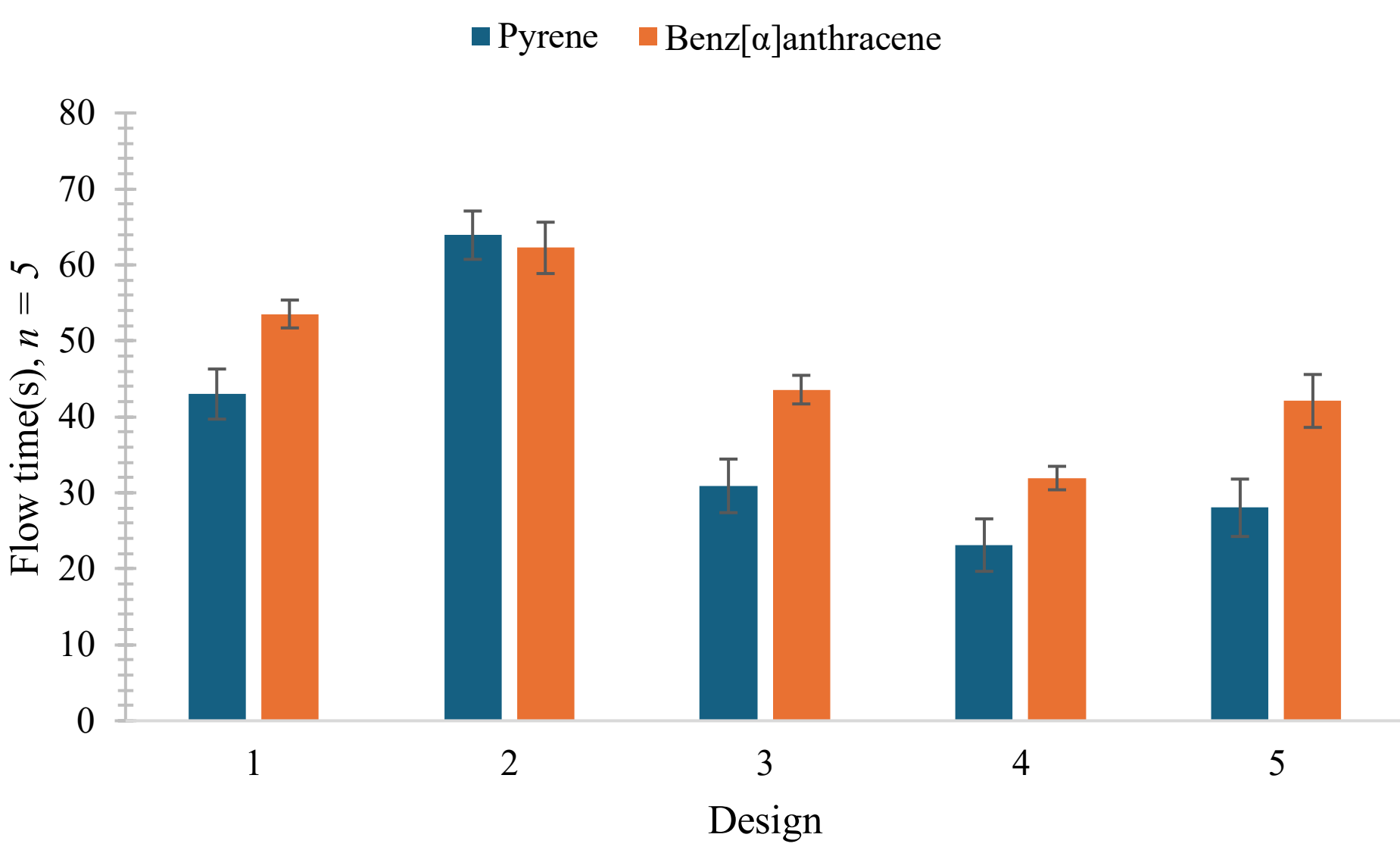

Figure 3. Flow transport time for pyrene and benz[α]anthracene across all five channel geometries.

Subsequently, CFD simulations were carried out to extract mean and peak velocities for comparison with experimental data, summarized in Table 2. Figure 4 shows the iso-surface velocity magnitude distributions across all five geometry designs. Peak velocity zones, visible as orange-to-red regions, appear at different spatial positions across designs, reflecting the unique flow behaviour imposed by each geometry. In Design 1, the peak velocity (0.099 mm/s) was observed in the converging zone, where channel narrowing accelerates the fluid, followed by deceleration in the subsequent diverging section; the remaining channel domain displays a moderate green-range velocity magnitude.

Design 2 exhibits two distinct high-velocity zones at its diverging and converging points (peak: 0.082 mm/s), capturing the alternating deceleration and acceleration process as the channel first widens then narrows. Design 3 shows concentrated high velocity at its narrow zone (peak: 0.089 mm/s), where the constriction sustains acceleration across the narrow section. In contrast, Designs 4 and 5 display elevated velocity magnitude upstream of their preconcentration zones before transitioning to lower velocity downstream. This behaviour is consistent with flow continuity: as fluid enters the wider preconcentration zone, cross-sectional area increases and velocity decreases to conserve mass flux, in accordance with the continuity equation for incompressible flow. The result is a characteristic high-velocity upstream region followed by a low-velocity accumulation zone, a pattern reported in radially expanding and uniform open-channel paper flow systems where geometric transitions govern local velocity redistribution. Design 4 recorded the highest peak velocity among all designs (0.104 mm/s), consistent with its circular geometry sustaining the broadest radial pressure distribution. The high red colour intensity in Design 4 CFD simulation also support the experimental data of observing the fastest flow transport among the design.

This geometry-driven variation is consistent with the predictions of the adapted LW equations (Eqs. 2–6). The standard LW equation (Eq. 1) describes capillary penetration in uniform channels, where flow velocity is governed by a constant capillary pressure term. The geometry-adapted forms introduce position-dependent terms that modify this pressure along the flow path. In Design 2, the diverging section progressively increases channel width, reducing capillary pressure and decelerating the fluid front (Eq. 3). Conversely, Design 4's radial expansion distributes capillary pressure uniformly across the advancing front, sustaining rapid wicking throughout the channel length (Eq. 5). For the intermediate designs, Design 3 (wide-to-narrow) drives elevated capillary pressure at the constriction, producing faster-than-average transport (Eq. 4), while Design 1 (converging-diverging) introduces competing zones of acceleration and retardation (Eq. 2), yielding a moderate net velocity. Design 5 (rectangular) shows near-uniform wicking with minimal geometric modulation (Eq. 6). The consistent ranking of transport times across both analytes confirms that these differences are governed by channel architecture rather than analyte-specific physicochemical properties. Benz[α]anthracene transported more slowly than pyrene across all designs, attributable to its higher molecular weight and extended aromatic structure, which increase molecular drag within the cellulose fiber matrix.

Table 2. Comparison of experimentally measured mean wetting-front velocity and CFD-extracted peak velocity and mean velocity for pyrene and benz[α]anthracene across all five channel geometries.

| Design | Channel Type | Exp. mean velocity (mm/s) | | CFD mean velocity | CFD Peak Velocity (mm/s) | Flow Behaviour |
|---|---|---|---|---|---|---|
| | | Pyrene | Benz[α]anthracene | | | |
| **1** | Converging-diverging | 0.449 | 0.344 | 0.536 | 0.099 | Moderate velocity gradient; converging zone accelerates flow, followed by deceleration in the subsequent diverging section |
| **2** | Diverging-converging | 0.284 | 0.284 | 0.382 | 0.082 | Low mean velocity; diverging section induces flow retardation |
| **3** | Wide-to-narrow | 0.558 | 0.422 | 0.591 | 0.089 | High mean velocity; constriction drives localized acceleration |
| **4** | Circular | 0.571 | 0.414 | 0.739 | 0.104 | High mean velocity; elevated velocity upstream of preconcentration zone transitions to lower velocity downstream via radial expansion |
| **5** | Rectangular | 0.480 | 0.355 | 0.505 | 0.073 | Near-uniform velocity; minimal geometric modulation |

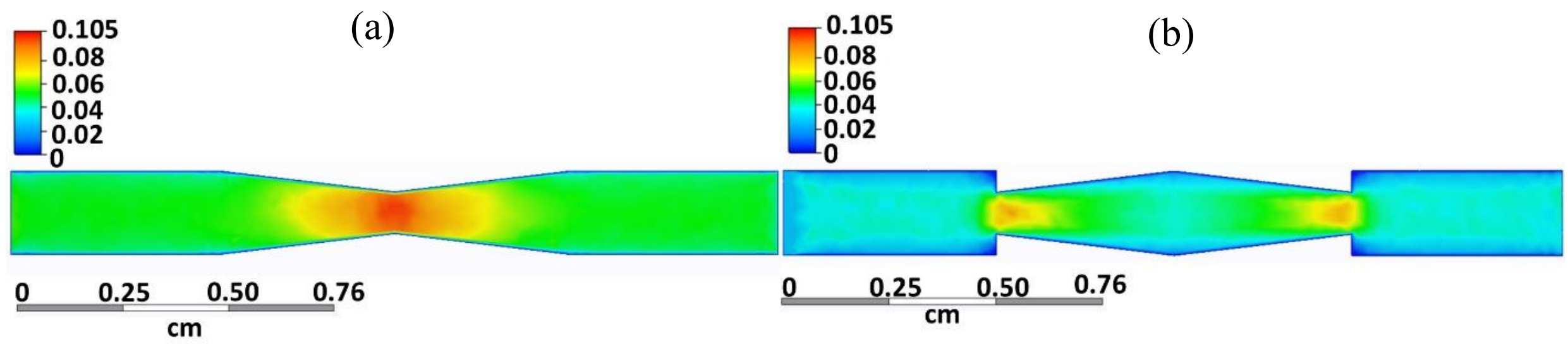

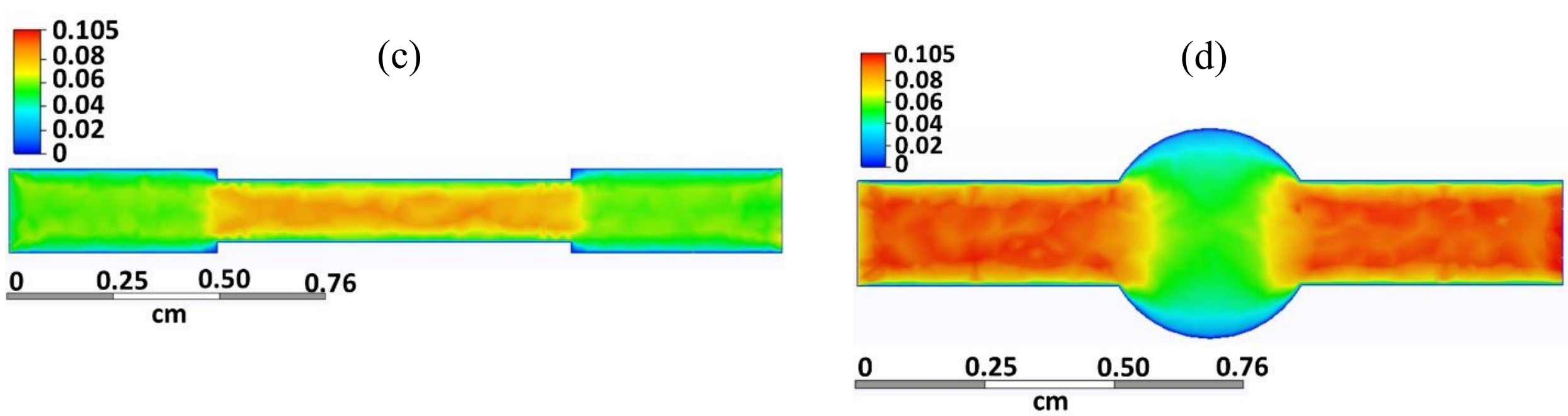

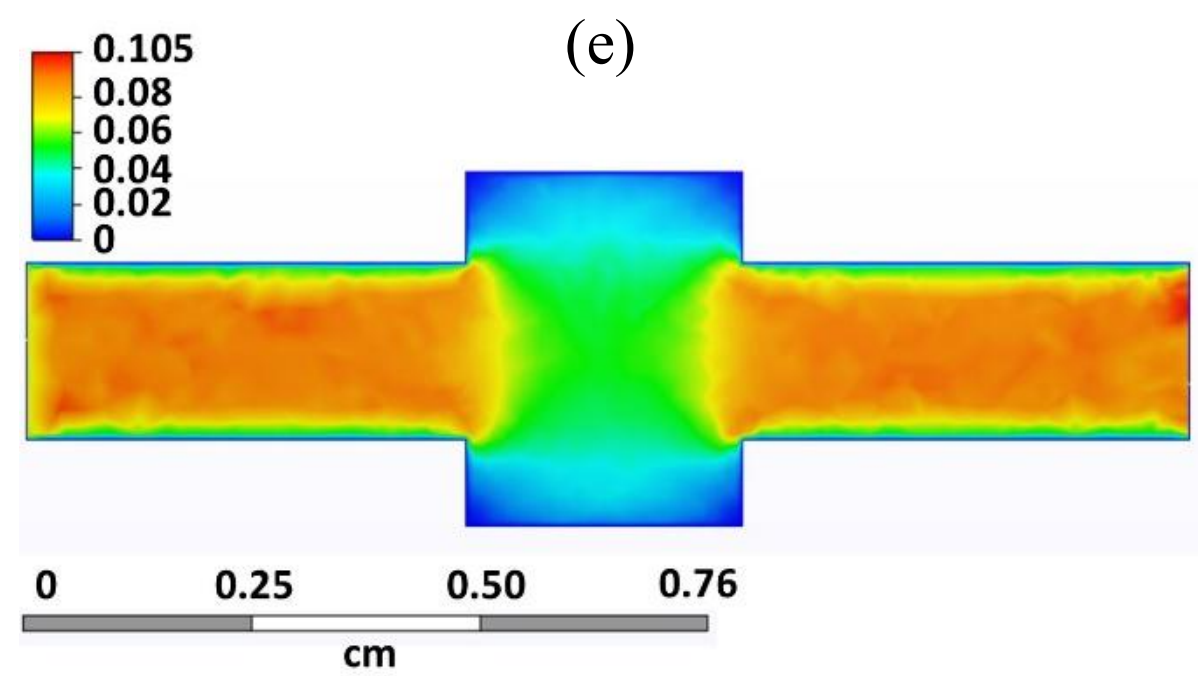

Figure 4. CFD velocity magnitude distributions for all five channel geometries: (a) Design 1: converging-diverging; (b) Design 2: diverging-converging; (c) Design 3: wide-to-narrow; (d) Design 4: circular; (e) Design 5: rectangular. High-velocity regions are shown in yellow-to-red; low-velocity regions in green-to-blue.

Evaporation has been recognised as a contributing factor to fluid transport behaviour in open paper-based microfluidic systems (Jeon et al. 2021). To evaluate its role in the present study, a gravimetric evaporation control study was conducted across all five geometries under identical fabrication and solvent conditions ($n = 5$). Evaporative mass loss varied across designs (Figure 5). All devices retained the majority of loaded solvent, and experiments were conducted under consistent ambient conditions, ensuring evaporation operated as a uniform background factor rather than a differential variable across geometries. The observed variation in evaporative loss across geometries, despite a consistent loaded volume and fixed preconcentration zone surface area of 32.5 $mm^2$, is suspected to be primarily attributable to differences in fluid transport time, which ranged from 23.1 s (Design 4) to 65.0 s (Design 2). As evaporation is a time-dependent process, longer transport durations may result in greater cumulative fluid exposure to ambient air, potentially contributing to higher mass loss in slower geometries (Jeon et al. 2021). The non-linear relationship between transport time and evaporative loss across all five designs further suggests that additional geometry-dependent factors, such as the spatial distribution of the fluid front and local vapour pressure gradients above the wetted surface, may also contribute to the observed variation (Cai et al. 2021; Berthier et al. 2016). Nevertheless, these differences therefore suggested to be a consequence of the geometry-driven capillary flow dynamics rather than independent confounding variables. While Gautam et al. (2025) demonstrated that fabrication technique can exert a pronounced influence on capillary imbibition, plotter cutting was classified in their study as a surface-intact technique producing the least substrate modification among all methods examined. Furthermore, a single fabrication method was applied consistently across all five designs in the present study, eliminating inter-fabrication variability as a confounding factor. Under these controlled conditions, channel geometry remains the primary variable governing the observed differences in flow behaviour and analyte confinement.

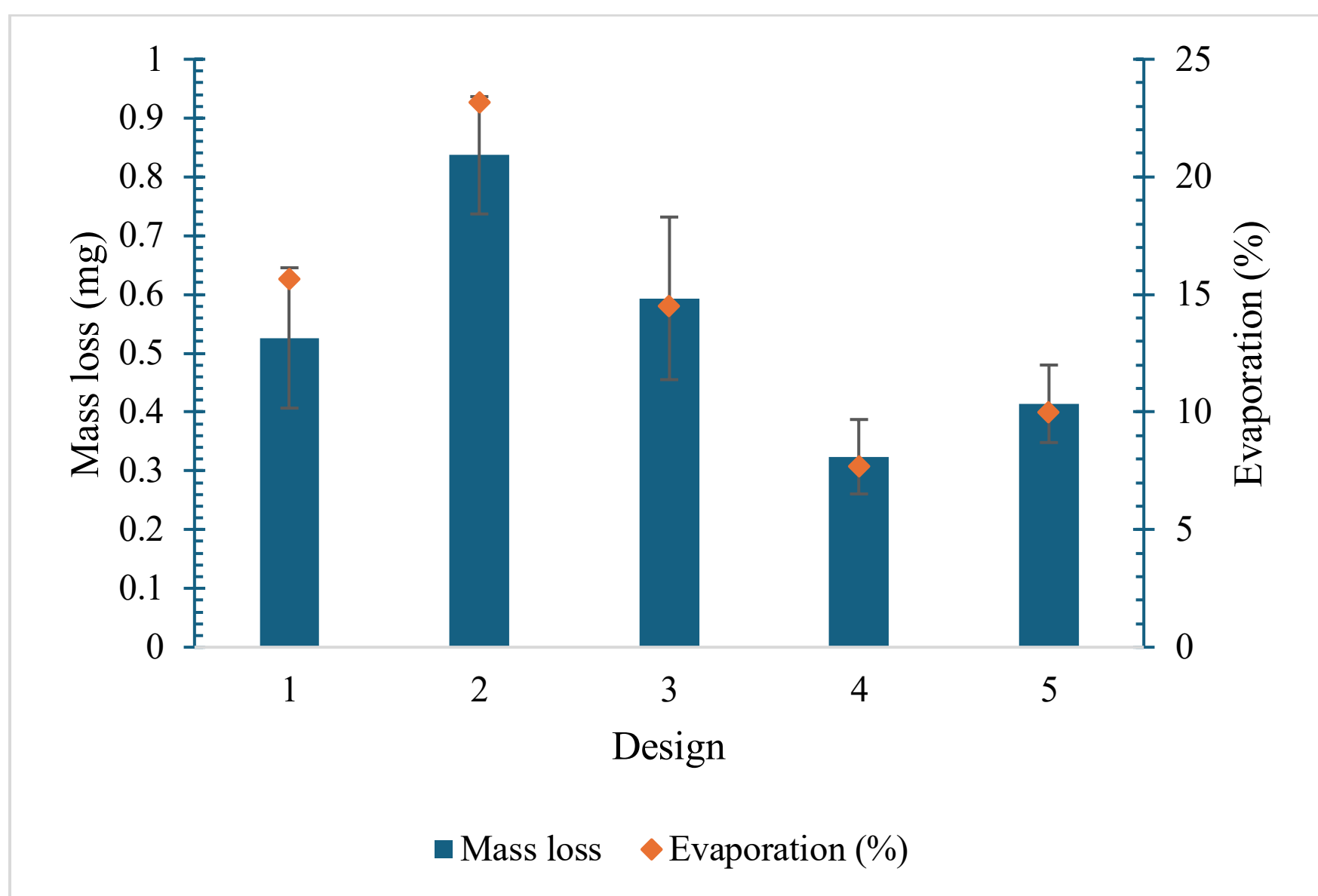

Figure 5 : Gravimetric evaporative mass loss across all five channel geometries, measured using methanol-loaded cyclodextrin-functionalized devices ($n = 5$). Mass loss expressed as mean ± standard error (mg) and as percentage of methanol mass lost by evaporation relative to the initial loaded mass ($m_1$).

*3.2. Geometric Influence on Analyte Confinement and Retention*

The fluorescence images obtained under UV illumination (Figure 6a) and the corresponding brightness quantification data (Figure 6b) reveal a clear dependence of analyte confinement on channel geometry. Devices functionalized with cyclodextrin-chitosan coatings consistently produced higher brightness values than non-functionalized controls across all geometries, confirming that the surface coating successfully anchors analyte accumulation at the preconcentration zone. The mean brightness values, control values, and percentage enhancements for all five designs are summarized in Table 3. The following analysis focuses on the differential geometric contribution, assessed by comparing brightness across designs relative to controls.

As depicted in Figure 7a, the highest confinement was observed in Design 2 (diverging-converging), which also exhibited the slowest flow transport time. The prolonged residence time in this geometry, driven by reduced capillary pressure in the diverging section (Eq. 3), provides an extended window for host-guest interaction between pyrene and the cyclodextrin surface coating, leading to greater accumulation. Quantitatively, Design 2 produced a mean brightness of 193.68, representing a 141.9% enhancement over its non-functionalized control (80.06) and a 54.3% higher brightness than Design 5 which was the weakest-performing geometry for pyrene (mean brightness: 125.49). The bright, spatially uniform fluorescence pattern observed in Design 2 replicates (Figure 6a) reflects this geometry-driven retention enhancement. In contrast, Designs 4 and 5 yielded only 11.5% and 6.5% enhancement over their respective controls, confirming that geometries lacking flow modulation provide negligible confinement benefit beyond the surface chemistry alone, consistent with the uniform velocity distributions predicted by Eqs. 5 and 6.

For benz[α]anthracene (Figure 7b), the highest geometric enhancement was observed in Design 3 (wide-to-narrow), which produced a 77.9% improvement over its non-functionalized control (mean brightness: 146.26 vs. 82.20). The localized velocity increase at the channel constriction (Eq. 4) efficiently concentrates the advancing analyte front, maximizing the probability of inclusion complex formation within the functionalized zone. Design 1 produced the highest absolute brightness for benz[α]anthracene (151.69), reflecting a 54.9% enhancement over its control, attributable to its combined constriction and expansion zones that both accelerate and retard flow sequentially. The larger and more intense fluorescence areas observed for benz[α]anthracene compared to pyrene across most geometries are consistent with its greater hydrophobicity and molecular size, which reduce transport velocity and increase interaction time with cellulose fibers and the cyclodextrin coating (Patel et al. 2020; Laguerre and Gall 2023). Design 4 showed the weakest response with only a 3.3% enhancement over control, and Design 5 produced only 14.1%, confirming that geometries lacking flow modulation features (Eqs. 5 and 6) are ineffective at focusing analyte confinement regardless of analyte type.

Table 3. Mean fluorescence brightness of cyclodextrin-functionalized (CD) and non-functionalized control devices, and percentage enhancement attributable to cyclodextrin surface chemistry, for pyrene and benz[α]anthracene across all five channel geometries.

| **Design** | **Pyrene CD** | **Pyrene Control** | **Pyrene Enh. (%)** | **BaA CD** | **BaA Control** | **BaA Enh. (%)** |
|---|---|---|---|---|---|---|
| **1** | 121.44 | 84.86 | +43.1 | 151.69 | 97.90 | +54.9 |
| **2** | **193.68** | 80.06 | **+141.9** | 145.22 | 114.91 | +26.4 |
| **3** | 173.63 | 108.69 | +59.8 | **146.26** | 82.20 | **+77.9** |
| **4** | 130.25 | 116.78 | +11.5 | 130.82 | 126.68 | +3.3 |
| **5** | 125.49 | 117.80 | +6.5 | 141.02 | 123.57 | +14.1 |

Bold values indicate the best-performing geometry per analyte. Enhancement (%) = [(CD − Control) / Control] × 100. BaA = benz[α]anthracene; CD = cyclodextrin-functionalized device.

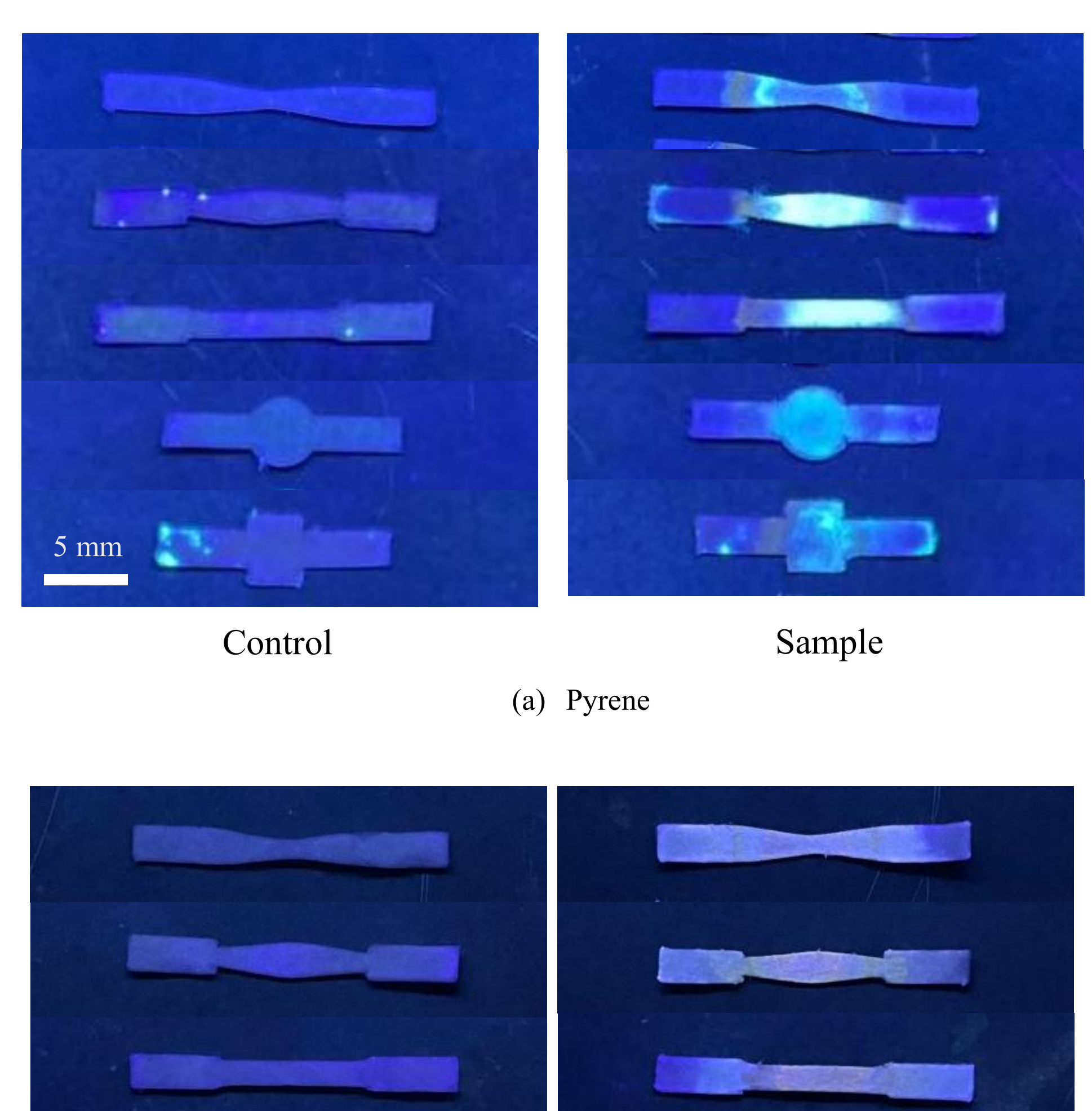

(a) Pyrene

5 mm
Control
Sample

(b) Benz[α]anthracene

Figure 6. UV fluorescence images of analyte confinement in preconcentration zones for (a) pyrene and (b) benz[α]anthracene: Control = non-functionalized control; Sample = replicate for each geometry.

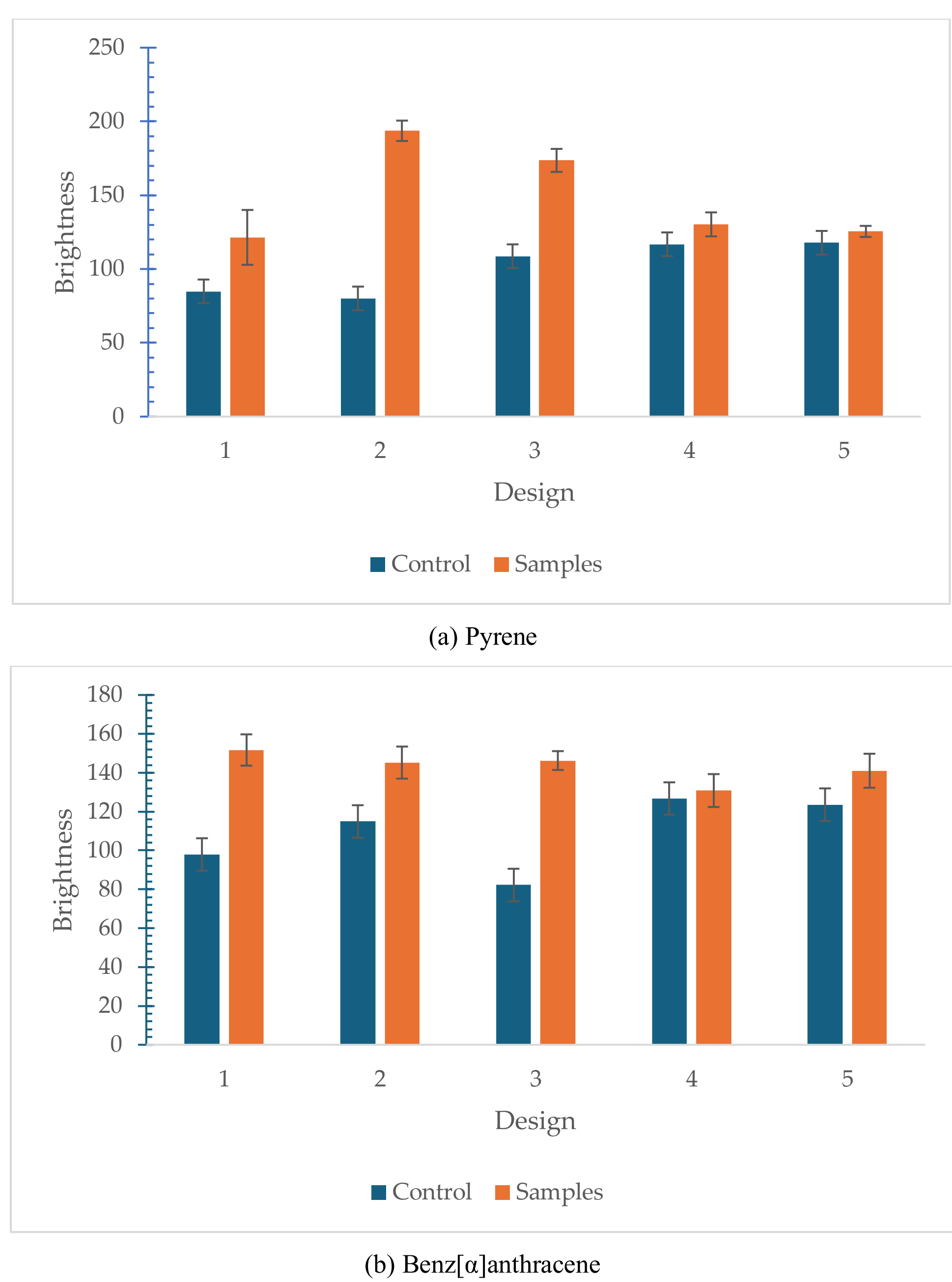

(a) Pyrene

(b) Benz[α]anthracene

Figure 7. Quantified fluorescence brightness of preconcentration zones for (a) pyrene and (b) benz[α]anthracene across all five channel designs. Error bars represent standard deviation across five replicates.

These findings reveal an important design consideration: optimal geometry is analyte dependent. Geometries that impose flow retardation (Design 2) favor analytes with higher mobility, while geometries that generate localized acceleration and constriction (Design 3) are more effective for larger, slower-moving analytes. For practical device design, this implies that geometry selection should be guided by the physicochemical properties of the target analyte. Future work could explore a universal geometry that provides near-optimal confinement across a range of analyte types, as well as the extension of these geometric design principles to other porous substrates and fluid systems.

Figure 8 presents the mean velocity difference between CFD and experimental results for both analytes across all five designs. For non-uniform geometries (Designs 1, 2, and 3), CFD consistently overpredicts experimental means for both analytes. The largest overprediction was observed in Design 3 (pyrene: +0.033 mm/s; benz[α]anthracene: +0.169 mm/s), driven by its near-full-channel high-velocity domain inflating the CFD domain average. Design 2 produced the smallest overprediction for both analytes (pyrene: +0.098 mm/s; benz[α]anthracene: +0.098 mm/s), consistent with its retardation-dominant character suppressing local velocity peaks. For uniform-cross-section geometries (Designs 4 and 5), CFD slightly underpredicts pyrene experimental means (Design 4: −0.040 mm/s; Design 5: −0.012 mm/s) while overpredicting benz[α]anthracene means (+0.117 and +0.113 mm/s respectively), reflecting greater pore-scale resistance experienced by the larger molecule as an effect the hollow-channel model cannot resolve. CFD correctly identifies Design 2 as the slowest and Design 4 as the fastest geometry, confirming that the simulation captures geometry-driven flow modulation even where absolute velocity agreement is limited by the simplified porous-medium representation. Benz[α]anthracene generally exhibited slower transport than pyrene across all geometries, attributable to its higher molecular weight and more extended aromatic structure, which increase molecular drag and interaction with cellulose fibers in the porous paper matrix (Patel et al. 2020). This analyte-dependent difference in transport rate further validates the use of two physicochemically distinct model compounds as probes of geometry-driven flow behavior.

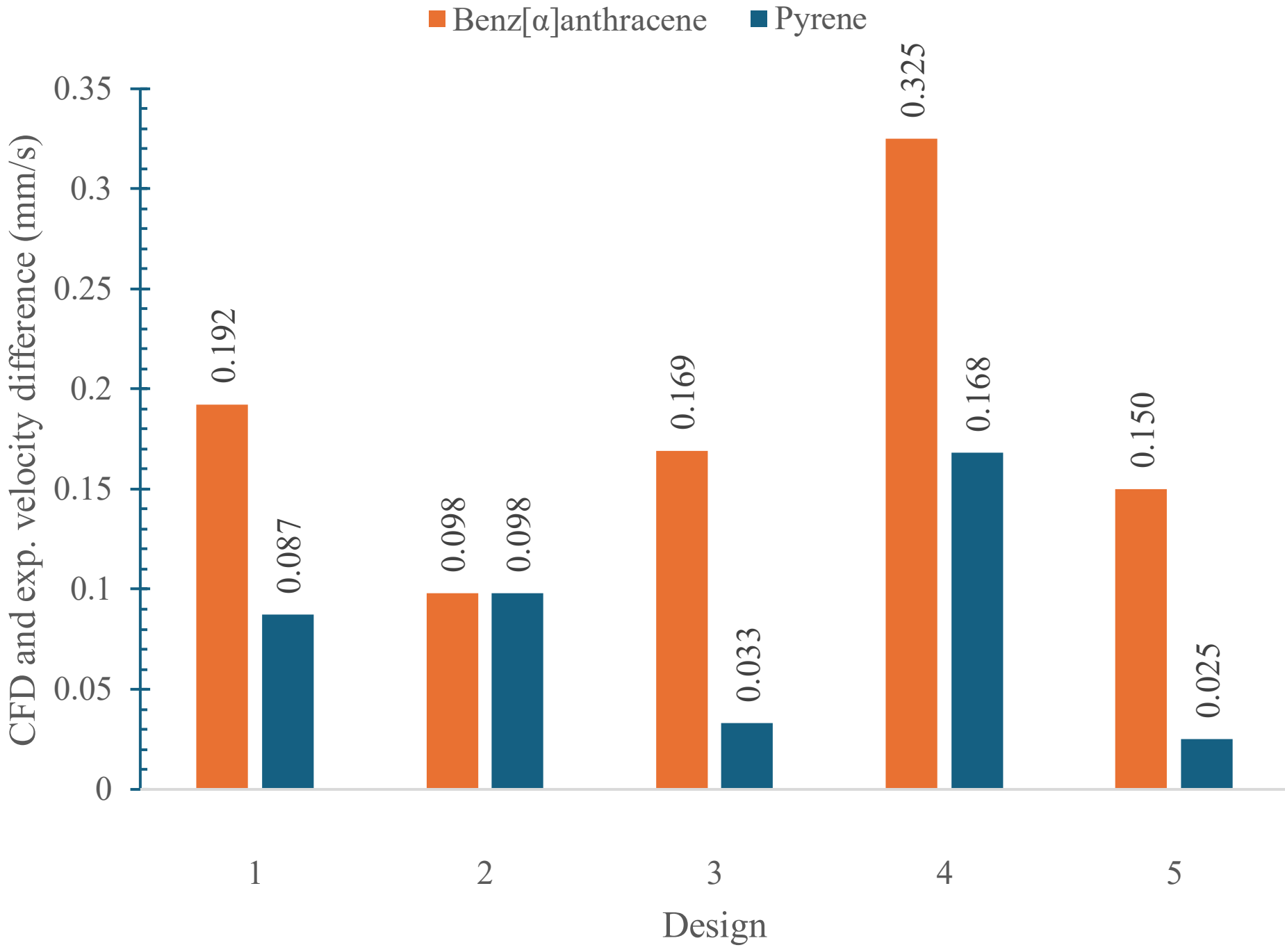

Figure 8. Experimental velocity difference and CFD for pyrene and benz[α]anthracene across all five channel geometries.

*3.3. CFD Mesh Independence and Simulation Validation*

Mesh independence was confirmed for all five designs as described in Section 2.5, with outlet velocity converging within 3% between coarse and medium mesh levels. The validated mesh configurations were used for all reported simulations.

The present section focuses on the quantitative comparison between CFD-derived velocities and experimental means, as summarized in Table 2. CFD mean velocities, extracted from the full channel domain, provide a more representative basis for comparison with experimental wetting-front means than peak velocities, which reflect only the maximum local value at high-gradient zones. This distinction is most evident in Design 2: despite recording the highest CFD peak velocity (0.082 mm/s), it produced the lowest CFD mean (0.382 mm/s) and experimental means for both analytes (pyrene: 0.284 mm/s; benz[α]anthracene: 0.284 mm/s), directly reflecting its retardation-dominant flow character.

For Designs 4 and 5, whose cross-sections are relatively uniform, CFD mean velocities showed close agreement with pyrene experimental means where Design 4 deviated by 7.0% (0.739 vs 0.571 mm/s) and Design 5 by 2.5% (0.505 vs 0.480 mm/s). Agreement was weaker for benz[α]anthracene, with deviations of 28.3% and 31.8% respectively, attributable to greater pore-scale analyte-matrix interaction of the larger molecule resulting as an effect not resolved by the hollow-channel CFD model. For non-uniform geometries (Designs 1, 2, and 3), CFD consistently overpredicts experimental means for both analytes, with Design 3 showing the largest deviation (pyrene: +0.033 mm/s; benz[α]anthracene: +0.169 mm/s), driven by its near-full-channel high-velocity domain inflating the domain average.

Overall, CFD correctly identifies Design 2 as the slowest and Design 4 as the fastest geometry. However, full ranking agreement across all five designs is not achieved, particularly between Designs 1 and 4, reflecting pore-scale phenomena beyond the resolution of the hollow-channel model. The simulation therefore serves as a qualitative and semi-quantitative tool for geometry-driven flow assessment, reliable for identifying flow behaviour trends and extreme-performing geometries, but not sufficient for precise absolute velocity prediction.

**4. Conclusions**

The present study demonstrates that channel geometry is an important determinant of capillary-driven flow behaviour and analyte confinement in paper-based devices. Five channel designs were fabricated and characterized experimentally, with transport times ranging from 23.1 s to 65.0 s driven entirely by geometric differences at constant substrate area. The geometry-adapted LW equations (Eqs. 2–6) successfully predicted the direction and relative magnitude of these differences, confirming that non-uniform capillary channels in porous media can be described by straightforward extensions of classical LW theory.

CFD simulations, validated through a mesh independence study, reproduced the experimentally observed velocity magnitude distributions across all five geometries and yielded CFD mean velocities that correctly identify Design 2 as the slowest and Design 4 as the fastest geometry. Semi-quantitative agreement was achieved for pyrene experimental means in Designs 4 and 5 (deviations of 7.0% and 2.5% respectively), while larger deviations for benz[α]anthracene reflect pore-scale analyte-matrix interactions not resolved by the model. The simulation serves as a reliable qualitative and semi-quantitative tool for geometry-driven flow assessment.

Analyte confinement, quantified by UV fluorescence brightness in cyclodextrin-functionalized preconcentration zones, showed a clear geometry dependence. Diverging-converging channels produced the highest confinement for pyrene by extending analyte residence time, while wide-to-narrow channels were most effective for benz[α]anthracene through sustained acceleration across the constricted section. Circular and rectangular geometries, which impose uniform velocity fields, were ineffective at focusing analyte confinement regardless of analyte type. The cyclodextrin-chitosan surface coating functioned as a retention anchor enabling spatial quantification of geometric effects, consistently enhancing brightness relative to non-functionalized controls.

Several limitations of the present study should be noted. First, the optimal geometry was found to be analyte-dependent, meaning that a single design may not provide optimal confinement for multiple analytes simultaneously. Future work should investigate broadly effective geometry designs that provide near-optimal performance across a range of analyte types. Second, the CFD model employed a simplified representation of the paper substrate; while this approach captures geometry-driven velocity gradients, it does not resolve pore-scale transport phenomena. Additionally, while the gravimetric evaporation control study suggested that differential evaporation does not account for the observed inter-geometry differences under the present experimental conditions, the robustness of these findings across varying environmental conditions such as humidity and temperature was not systematically evaluated and represents a direction for future investigation. Finally, the analyte concentrations used were selected to ensure adequate fluorescence signal for geometric comparison and do not represent detection limits or operational analytical ranges. Extension of these geometric design principles to lower concentration regimes and integration with downstream detection modules are important directions for future investigation.

The geometry-based design framework established here provides practical guidance for engineers developing passive paper-based devices requiring controlled fluid transport, spatially defined analyte retention, or tunable residence times. The principles are not specific to PAH analytes and are expected to generalize to other hydrophobic or surface-interactive compounds in porous-media channel systems.

Author Contributions: Conceptualization, M.S.N., A.K., and M.S.A.F.; investigation, M.S.N.; writing—original draft preparation, M.S.N.; writing—review and editing, A.K. and M.S.A.F.; supervision, A.K. and M.S.A.F.; funding acquisition, A.K. All authors have read and agreed to the published version of the manuscript.

Funding: This work was partly supported by JSPS KAKENHI (Grant Number 25K08827), and the Satake Technical Foundation.

Conflicts of Interest: The authors declare no conflicts of interest.

**Abbreviations**

The following abbreviations are used in this manuscript:

| | |
|---|---|
| **µPADs** | Paper-Based Microdevices |
| **PAH** | Polycyclic Aromatic Hydrocarbon |
| **CFD** | Computational Fluid Dynamics |
| **LW** | Lucas–Washburn |
| **RGB** | Red, Green, Blue |
| **SDS** | Sodium Dodecyl Sulfate |